\documentclass[a4paper,11pt]{article}
\usepackage{mathbbold}%
\usepackage{amsfonts}
\usepackage{bbding}
\usepackage{amssymb}
\usepackage{mathrsfs}
\usepackage{float}
\usepackage{amsmath}
\usepackage{amsfonts,amssymb}
\usepackage{graphicx,color}
\usepackage{psfig}
\usepackage{mathrsfs}
\usepackage{graphicx}%
 \catcode`\@=11 \catcode`\@=12

\catcode`\@=11 \@addtoreset{equation}{section}
\def\theequation{\thesection.\arabic{equation}}
\catcode`\@=12 
\title{A new kind of solutions of the Einstein¡¯s field equations with
non-vanishing cosmological constant}
\author{De-Xing Kong\thanks{Department of Mathematics,
Zhejiang University, Hangzhou 310027, China;},$\quad$ Kefeng
Liu\thanks{Department of Mathematics, University of California at
Los Angeles, CA 90095, USA;} $\quad$ and $\quad$ Ming
Shen\thanks{Center of Mathematical Sciences, Zhejiang University,
Hangzhou 310027, China.}}
\date{ }
\begin{document}
\maketitle
\begin{abstract}
In this paper we construct a new kind of solutions of the
Einstein¡¯s field equations with non-vanishing cosmological
constant, which possess some interesting physical properties. The
singularities of this kind of solutions are investigated. According
to the general form of this kind of solutions, we give some
examples.

\noindent\textbf{Key words and phrases}:Einstein¡¯s field equations,
cosmological constant, Riemann curvature tensor, Weyl scalars,
singularity.

\vskip 3mm

\noindent\textbf{2000 Mathematics Subject Classification}:
04.20.Jb; 98.80.Jk; 02.30.Jr
\end{abstract}

\newpage\baselineskip=6mm

\section{Introduction}
 The Einstein¡¯s field equations with cosmological constant take the form
\begin{equation}
R_{\mu\nu}-\frac{1}{2}g_{\mu\nu}R+\Lambda g_{\mu\nu}=\frac{8 \pi
G}{c^4}T_{\mu\nu}.
\end{equation}
where $g_{\mu\nu}(\mu,\nu  = 0, 1, 2, 3)$ is the unknown Lorentzian
metric, $R_{\mu\nu}$ is the Ricci curvature tensor, $R
=g^{\mu\nu}R_{\mu\nu}$ is the scalar curvature in which $g^{\mu\nu}$
is the inverse of $g_{\mu\nu}$, $\Lambda$ is the cosmological
constant, G stands for the Newton¡¯s gravitational constant, c is
the velocity of the light and $T_{\mu\nu}$ is the energy-momentum
tensor. In a vacuum, i.e., in regions of space-time in which
$T_{\mu\nu}=0$, the Einstein¡¯s field equations (1.1) reduce to
\begin{equation}
R_{\mu\nu}-\frac{1}{2}g_{\mu\nu}R+\Lambda g_{\mu\nu}=0,
\end{equation}
or equivalently,
\begin{equation}
R_{\mu\nu}=\Lambda g_{\mu\nu}.
\end{equation}

 The study on exact
solutions of the Einstein's field equations with cosmological
constant has a long history. There have been many beautiful results
on this research topic (e.g. \cite{a1}, \cite{a2}, \cite{a3}). As
early as 1918, Kottler extended the Schwarzschild solution and
obtained the static spherically symmetric exterior solution with
cosmological constant (see \cite{b}). In 1951, Nariai obtained a
spherical symmetric and static character cosmological solution which
could not be transformed into the standard form(see \cite{c}).
Kramer et al got the Reissner-Nordsr\"{o}m exterior solution with
cosmological constant (see \cite{d}) in 1980. Xu
 et al (\cite{e}) extended the Florides' solution to the
case with cosmological constant in 1987. Recently, Kong and Liu
presented a metric of the following form
\begin{equation}\left(g_{\mu\nu}\right)=\left( \begin{array}{cccc}
u & v & p & q \\
v & w & 0 & 0 \\
p & 0 & \rho & 0 \\
q & 0 & 0 & \sigma \\
\end{array}\right),\end{equation}
where $u,v,p,q,w,\rho$ and $\sigma$ are smooth functions of the
coordinates $(t,x,y,z)$, and they proved that under the assumption
\begin{equation}g\stackrel{\triangle}{=}\det(g_{\mu\nu})=
uw\rho\,\sigma-{v}^{2}\rho\,\sigma-{p}^{2}w\sigma-{q}^{2}w\rho\ <0
\end{equation} the
metric $(g_{\mu\nu})$ is Lorentzian, and they constructed  some new
solutions of the vacuum Einstein's field equations with vanishing
cosmological constant (see \cite{f}).

In cosmology, the cosmological constant $\Lambda$ was proposed by
Einstein as a modification of his original theory of general
relativity to achieve a stationary universe. Unfortunately, Einstein
abandoned the concept after the observation of the Hubble redshift
indicated that the universe might not be stationary, since he had
based his theory off the idea that the universe is unchanging (see
\cite{g}). However, the discovery of cosmic acceleration in the
1990s has renewed the interest in a cosmological constant. The
Einstein field equation with cosmological constant becomes one of
the hot research problems (\cite{g1}-\cite{g3}).

In this Paper, we construct a kind of solutions of the vacuum
Einstein's field equations with non-vanishing cosmological constant.
We show that the Riemann curvature tensor and the norm of the
Riemann curvature tensor do not disappear when $ \Lambda\neq 0 $.
According to the general form of this kind of solutions, we give
some examples.

\section{ Solutions with cosmological constant $ \Lambda $ }

In this section, we first construct a new kind of solutions to the
equations (1.3), then calculate the Riemann curvature tensors, the
norm of the Riemann curvature tensors and then analyze the Weyl
scalars of the solutions constructed.

In the coordinates $(t,x, y,z)$, we consider the following line
element
\begin{equation}
ds^2=(dt,dx,dy,dz)(g_{\mu\nu})(dt,dx,dy,dz)^T=udt^2+2vdtdz-a^2b^2dx^2-a^2dy^2,
\end{equation}
where $u$, $v$, $a$ are smooth functions of $t$, $x$, and $b$ is a
smooth function of $t$. It is easy to verify that the determinant of
$(g_{\mu\nu})$ is given by
\begin{equation*}
g\stackrel{\triangle}{=}\det(g_{\mu\nu})=-v^2a^4b^2.
\end{equation*}
Noting (1.5), it is easy to observe that the metric $(g_{\mu\nu})$
is Lorentzian.

Using the line element (2.1), the field equations (1.3) are reduced
to the following system of differential equations
\begin{equation}
\frac{uv_x^2+u_{xx}v^2-u_xv_xv}{2}+vv_ta^2bb_t-v^2a^2bb_{tt}-2v^2aba_tb_t-2v^2ab^2a_{tt}+2vab^2v_ta_t=\Lambda
a^2b^2v^2u,
\end{equation}
\begin{equation}
v^2ba_xa_t+v^2aa_xb_t+vaba_tv_x-v^2aba_{xt}+\frac{va^2v_xb_t+a^2bv_tv_x-a^2bvv_{xt}}{2}=0,
\end{equation}
\begin{equation}
v_{xx}=2\Lambda a^2b^2v,
\end{equation}
\begin{equation}
2v_{xx}va^2-2avv_xa_x-v_x^2a^2-2v^2a_x^2+2v^2aa_{xx}=2\Lambda
a^4v^2b^2,
\end{equation}
\begin{equation}
av_xa_x+vaa_{xx}-va_x^2=\Lambda a^4b^2v.
\end{equation}
Here we only consider the case that the cosmological constant
$\Lambda>0$. In order to simplify the above system, we take the
following ansatz
\begin{equation}
a^2b^2=\frac{3}{\Lambda x^2}.
\end{equation}
Substituting (2.7) into (2.4) yields
\begin{equation}
v_{xx}-\frac{6v}{x^2}=0.
\end{equation}
Equation (2.8) gives
\begin{equation}
v=mx^3+\frac{c}{x^2},
\end{equation}
where $m=m(t)$ and $c=c(t)$ are integral functions. Substituting
(2.7) and (2.9) into (2.6), it is easily to get
\begin{equation*}
m=0,
\end{equation*}
so the equation (2.9) becomes
\begin{equation}
v=\frac{c}{x^2}.
\end{equation}
When $a^2b^2$ and $v$ are given by (2.7) and (2.10), the equations
(2.3) and (2.5) are automatically satisfied. And the equation (2.2)
is simplified to
\begin{equation}
x^2u_{xx}+2xu_x-2u=\frac{6}{\Lambda}(\frac{2b_t^2}{b^2}-\frac{b_{tt}}{b}+\frac{b_tc_t}{bc}).
\end{equation}
Solving (2.11), we obtain
\begin{equation}
u=\frac{3}{\Lambda}(\frac{f}{x^2}+hx-\frac{2b_t^2}{b^2}+\frac{b_{tt}}{b}-\frac{b_tc_t}{bc}),
\end{equation}
where $f$ and $h$ are integral functions of $t$.

From the above discussion, it is easy to have Theorem 1.

 \noindent{\bf Theorem 1}{ The vacuum
Einstein's field equations with cosmological constant (1.3) have the
following family of solutions in the coordinates $(t,x,y,z)$
\begin{equation}
ds^2=\frac{3}{\Lambda}(\frac{f}{x^2}+hx-\frac{2b_t^2}{b^2}+\frac{b_{tt}}{b}-\frac{b_tc_t}{bc})dt^2+\frac{2c}{x^2}dtdz-\frac{3}{\Lambda
x^2}dx^2-\frac{3}{\Lambda x^2b^2}dy^2,
\end{equation}
where $f$, $h$, $c$ are integral functions of $t$, and $b$ is an
arbitrary smooth function of t.
}$\qquad\qquad\qquad\qquad\qquad\qquad\qquad\qquad\qquad \square$

  By direct calculations, the scalar curvature of (2.13) equals to
$4\Lambda$, and the Riemann curvature tensor of (2.13) reads

\begin{equation}
R_{0101}=\frac{6fb^2c+15hx^3b^2c+6b_{tt}x^2bc-12b_t^2x^2c-6b_tc_tx^2b}{2\Lambda
x^4b^2c},
\end{equation}
\begin{equation}
R_{0202}=\frac{6fb^2c-3hx^3b^2c+6b_{tt}x^2bc-12b_t^2x^2c-6b_tc_tx^2b}{2\Lambda
x^4b^4c},
\end{equation}
\begin{equation}
R_{0303}=\frac{\Lambda c^2}{3x^4},
\end{equation}
\begin{equation}
R_{0113}=-\frac{c}{x^4},
\end{equation}
\begin{equation}
R_{0223}=-\frac{c}{x^4b^2},
\end{equation}
\begin{equation}
R_{1212}=-\frac{3}{\Lambda x^4b^2}
\end{equation}
and the other $R_{\alpha\beta\mu\nu}=0$. Moreover, we get the norm
of the Riemann curvature tensors
\begin{equation}\mathbf{R}\triangleq R^{ijkl}R_{ijkl}=\frac{8\Lambda^2}{3}.
\end{equation}

In general relativity, the Weyl scalars are a set of five complex
scalar quantities $\Psi_0, ..., \Psi_4$ describing the curvature of
a four-dimensional space-time. They are the expressions of the ten
independent degrees of freedom of the Weyl tensor
$C_{\mu\nu\alpha\beta}$ in the Newman-Penrose formalism. According
to the physical interpretation in Szekeres \cite{i}, $\Psi_0$ and
$\Psi_3$ are ingoing and outgoing longitudinal radiation terms while
$\Psi_1$ and $\Psi_4$ are ingoing and outgoing transverse radiation
terms, and $\Psi_2$ is a $Coulomb$ $term$ representing the
gravitational monopole of the source. By GRTENSOR II program, the
Weyl scalars of (2.13) read
\begin{equation}
\Psi_0=\Psi_1=\Psi_2=\Psi_3=0,
\end{equation}
\begin{equation}
\Psi_4=-\frac{3x^5h}{2c^2}.
\end{equation}

  From the above calculations, we easily have the following
  property.

\noindent{\bf Property 1}~~ {For any fixed $t \in \mathbb{R}$, the
non-vanishing $ R_{\alpha\beta\mu\nu} \longrightarrow \infty$ as
$x\longrightarrow 0$.}

On the other hand, the norm of the Riemann curvature tensors
$\mathbf{R}=\displaystyle{\frac{8\Lambda^2}{3}}$ is invariable. This
suggests that $x=0$ is not a real physical singularity but a
geometric one which is a result of a bad choice of coordinate.

\section{Examples}

In this section, we give some examples according to (2.13) in
Theorem 1 and give some detailed analysis.

 \noindent{\bf  Example 1}

 Let $f=h=t^2+1$ and $c=b=1$.
The line element (2.13) is reduced to
\begin{equation}
ds^2=\frac{3(t^2+1)}{\Lambda}(\frac{1}{x^2}+x)dt^2+\frac{2
}{x^2}dtdz-\frac{3}{\Lambda x^2}dx^2-\frac{3}{\Lambda x^2}dy^2.
\end{equation}
It is easy to get
\begin{equation}
g\stackrel{\triangle}{=}\det(g_{\mu\nu})=-\frac{9}{\Lambda^2x^8}.
\end{equation}
According to (2.14)-(2.19), the non-zero components of Riemann
curvature tensor of (3.1) are
\begin{equation}
R_{0101}=\frac{(t^2+1)(6+15x^3)}{2\Lambda x^4},
\end{equation}
\begin{equation}
R_{0202}=\frac{(t^2+1)(6-3x^3)}{2\Lambda x^4},
\end{equation}
\begin{equation}
R_{0303}=\frac{\Lambda }{3x^4},
\end{equation}
\begin{equation}
R_{0113}=-\frac{1}{x^4},
\end{equation}
\begin{equation}
R_{0223}=-\frac{1}{x^4b^2},
\end{equation}
\begin{equation}
R_{1212}=-\frac{3}{\Lambda x^4}.
\end{equation}
From (3.3)-(3.8), we have the corresponding Riemann curvature tensor
satisfies Property 1. Thus $x=0$ is a geometric singularity of the
spacetime described by (3.1).

Moreover, it is easy to see that the t-axis stands for the time-axis
when $ x> -1$ and the z-axis stands for the time-axis when $x< -1$.
When $x=-1$, the line element (3.1) reduce to
\begin{equation}
ds^2=2dtdz-\frac{3}{\Lambda }dy^2.
\end{equation}
When $x>-1$, for any fixed $t\in \mathbb{R}$, it follows from (3.1)
that the induced metric of the $time$-slice reads
\begin{equation}
ds^2=-\frac{3}{\Lambda x^2}(dx^2+dy^2).
\end{equation}
When $x<-1$,  for any fixed $z\in \mathbb{R}$, the induced metric of
the $time$-slice is
\begin{equation}
ds^2=\frac{3(t^2+1)}{\Lambda}(\frac{1}{x^2}+x)dt^2-\frac{3}{\Lambda
x^2}(dx^2+dy^2).
\end{equation}

\noindent{\bf  Example 2}

 Taking $f=2+\sin t$, $c=\cos t e^{-\sin
t}$, $h=0$ and $b=e^{\sin t}$, the line element (2.13) becomes
\begin{equation}
ds^2=\frac{3(2+\sin t)}{\Lambda x^2}dt^2+\frac{2\cos t }{x^2e^{\sin
t}}dtdz-\frac{3}{\Lambda x^2}dx^2-\frac{3}{\Lambda x^2e^{2\sin
t}}dy^2.
\end{equation}
In this case, we have
\begin{equation}
g\stackrel{\triangle}{=}\det(g_{\mu\nu})=-\frac{9\cos^2
t}{\Lambda^2x^8e^{4\sin t}}.
\end{equation}
According to (2.14)-(2.19), the non-zero components of Riemann
curvature tensor of (3.12) are
\begin{equation}
R_{0101}=\frac{3(2+\sin t)}{\Lambda x^4},
\end{equation}
\begin{equation}
R_{0202}=\frac{3(2+\sin t)}{\Lambda x^4 e^{2\sin t}},
\end{equation}
\begin{equation}
R_{0303}=\frac{\Lambda \cos^2 t}{3x^4 e^{2\sin t}},
\end{equation}
\begin{equation}
R_{0113}=-\frac{\cos t }{x^4e^{\sin t}},
\end{equation}
\begin{equation}
R_{0223}=-\frac{\cos t }{x^4e^{4 \sin t}},
\end{equation}
\begin{equation}
R_{1212}=-\frac{3}{\Lambda x^4e^{2\sin t}}
\end{equation}
 From (3.13 ), we obtain that $t=k\pi+\pi/2$  $k\in \mathbb{N}$ is the
 degenerate singularity. Moreover, from (3.14)-(3.19), we get the Riemann curvature tensor of (3.12)
satisfies Property 1. Thus, we also have  $x=0$ is a geometric
singularity of the spacetime described by (3.12).

Fixing $y$ and $z$, we get the induced metric
\begin{equation}ds^2=\frac{3(2+\sin t)}{\Lambda x^2}dt^2-\frac{3}{\Lambda x^2}dx^2.\end{equation}
 Consider the null curves in
the $(t,x)$-plan, which are defined by
\begin{equation}\frac{3(2+\sin t)}{\Lambda x^2}dt^2-\frac{3}{\Lambda x^2}dx^2=0.\end{equation} Then we get
\begin{equation} \frac{dt}{dx}=\pm\frac{1}{\sqrt{2+\sin t}}\end{equation}
For any fixed $t\in \mathbb{R}$, it follows from (3.12) that the
induced metric of the $t$-slice reads
\begin{equation}
ds^2=-\frac{3}{\Lambda x^2}dx^2-\frac{3}{\Lambda x^2e^{2\sin
t}}dy^2.
\end{equation}

\noindent{\bf  Example 3}

 Take
 \begin{equation}
 f={\rm{sn}^2}(t,k),
 ~~ h={\rm{cn}^2}(t,k)\end{equation}
and
 \begin{equation}
  c={\rm{dn}^2}(t,k),~~b=\sqrt{3},\end{equation}
 where ${\rm{sn}}(t,k)$, ${\rm{cn}}(t,k)$ and ${\rm{dn}}(t,k)$ are Jacobi
elliptic functions in which $0<k^2<1$ is the elliptic modulus. Then
we can get the following solution
\begin{equation}
ds^2=\frac{3}{\Lambda}\Big(\frac{{\rm{sn}^2}(t,k)}{x^2}+{\rm{cn}^2}(t,k)x\Big)dt^2+\frac{2{\rm{dn}^2}(t,k)}{x^2}dtdz-\frac{3}{\Lambda
x^2}dx^2-\frac{1}{\Lambda x^2}dy^2.
\end{equation}
Let $k \rightarrow 0$, we get
\begin{equation}
ds^2=\frac{3}{\Lambda}\Big(\frac{\sin^2 t}{x^2}+x\cos^2 t
\Big)dt^2+\frac{2}{x^2}dtdz-\frac{3}{\Lambda
x^2}dx^2-\frac{1}{\Lambda x^2}dy^2.
\end{equation}
On the other hand, let $k \rightarrow 1$, then we have
\begin{equation}
ds^2=\frac{3}{\Lambda}\Big(\frac{\tanh^2 t}{x^2}+x {\rm sech}^2 t
\Big)dt^2+\frac{2{\rm sech}^2 t}{x^2}dtdz-\frac{3}{\Lambda
x^2}dx^2-\frac{1}{\Lambda x^2}dy^2.
\end{equation}
Here we do not give the detailed analysis of the solution (3.26)
which is similar to example 2.
 \section{Conclusion}
The general form (2.13) is a new family solutions of the vacuum
Einstein's field equations with non-vanishing cosmological constant.
According to it, we construct some new exact solutions and
investigate the properties enjoyed by these solutions. Moreover, by
the line element (2.13), we can also get some known solutions. For
example, take
\begin{equation} h=0,~~b=1,\end{equation} and
\begin{equation} c=\frac{3f}{2\Lambda},\end{equation} then we have
\begin{equation}
ds^2=\frac{3f}{\Lambda
x^2}dt^2+\frac{3f}{x^2\Lambda}dtdz-\frac{3}{\Lambda
x^2}dx^2-\frac{3}{\Lambda x^2}dy^2,
\end{equation}
Define \begin{equation}z=\tilde{z}-t\end{equation} and
\begin{equation}d\tilde{t}=fdt.\end{equation}
 In the coordinates $(\tilde{t}, x,
y, \tilde{z})$, the line element (4.3) becomes
\begin{equation}
ds^2=\frac{3}{\Lambda x^2}(d\tilde{t}d\tilde{z}-dx^2-dy^2),
\end{equation}
Next, define
\begin{equation}\tilde{t}=\bar{t}+\bar{z}\end{equation} and
\begin{equation}\tilde{z}=\bar{t}+\bar{z}.\end{equation}
Then in the coordinates $(\bar{t}, x, y, \bar{z})$, we obtain
\begin{equation}
ds^2=\frac{3}{\Lambda x^2}(d\bar{t}^2-d\bar{z}^2-dx^2-dy^2),
\end{equation}
which is  conformal to the Minkowski space time. Thus, the general
form (2.13) obtained in this paper is worth studying. We expect some
applications of this new kind of solutions with cosmological
constant presented in this paper to modern cosmology and general
relativity.

\end{document}